\documentclass[]{article}
\usepackage[round,longnamesfirst]{natbib}
\usepackage{csquotes,amsmath, amsthm, amssymb, amsbsy, mdwlist}

\usepackage[belowskip=-5pt,aboveskip=0pt]{caption}
\usepackage[hyperfigures=true]{hyperref}

\hypersetup{
pdfauthor = {},
pdftitle = {Spatial Temporal Exponential-Family Point Process Models},
pdfkeywords = {STEPP, social network analysis, spatial-temporal, point process, longitudinal, latent space, Markov, substance use.}
pdfsubject = {Modeling social dynamics}
}

\usepackage{ifthen}
\usepackage[usenames]{color}
\newcounter{content}
\newcounter{notes}
\newcounter{techreport}

\usepackage{amssymb,amsmath,amsthm}
\usepackage{graphicx}
\usepackage{mathrsfs} 
\usepackage{amsmath}
\usepackage{amssymb}
\usepackage{enumerate}
\usepackage[all]{xy}
\usepackage{color, graphics}
\usepackage{palatino, url, multicol}
\usepackage{subfigure}
\usepackage{mathrsfs}
\usepackage{multicol}
\usepackage{tabularx}

\newcommand\norm[1]{\left\lVert#1\right\rVert}

\theoremstyle{plain}

\theoremstyle{definition}

\usepackage{graphicx}
\usepackage{float}
\usepackage{rotating}
\usepackage{natbib}
\usepackage{color}
\usepackage{hypernat}
\usepackage{pdflscape}

\usepackage{graphics}
\usepackage{subfigure}
\usepackage[all]{xy}
\usepackage{amsmath}
\usepackage{amssymb}
\usepackage[all]{xy}
\usepackage{palatino, url, multicol}
\usepackage{mathrsfs}
\usepackage{multicol}
\usepackage{enumerate}

\usepackage{fullpage}
\usepackage{ifthen}

\title{Spatial Temporal Exponential-Family Point Process Models for the Evolution of Social Systems}

\author{Joshua D. EmBree\footnote{Adjunct Researcher, RAND Corporation, Santa Monica, CA 90401 (E-mail: \emph{jembree@rand.org}).} \and Mark S. Handcock\footnote{Professor of Statistics, Department of Statistics, University of California, Los Angeles, CA 90095-1554 (E-mail: \emph{handcock@ucla.edu}).}}

\begin{document}

\maketitle

\thispagestyle{empty}
\setcounter{page}{0}

\begin{abstract}
We develop a class of exponential-family point processes based on a latent social space to model the coevolution of social structure and behavior over time. Temporal dynamics are modeled as a discrete Markov process specified through individual transition distributions for each actor in the system at a given time. We prove that these distributions have an analytic closed form under certain conditions and use the result to develop likelihood-based inference. We provide a computational framework to enable both simulation and inference in practice. Finally, we demonstrate the value of these models by analyzing alcohol and drug use over time in the context of adolescent friendship networks. 
\end{abstract}
\vspace*{.3in}

\section{Introduction}

Social systems play a fundamental role in the dynamics of human behavior and interest in studying these systems is growing. For example, \citet{fujimoto2012social} investigate contagion mechanisms for the transmission of drinking and smoking behaviors through adolescent social networks.  However, work of this nature is often limited by a lack of realistic stochastic models for the phenomena of interest. For such models to be applicable, they must adequately represent the complexity of social relations and behavior as they coevolve over time. 

Most often, social relations are measured with dyadic tie variables, for example friendship, and then assembled to form networks. There are numerous stochastic models for the evolution of social networks. \citet{holland1977dynamic} provide one of the earliest continuous-time Markov models for the process by which social structure affects individual behavior.  Arguably, the most popular subclass of these continuous-time Markov models is the so called {\it stochastic actor-oriented model} (SOAM) described in \citet{snijders2005models} and \citet{snijders2010introduction} which are framed in the context of individual actors making decisions to form or break ties with other actors. \citet{snijders2007modeling} extend the SAOMs to jointly model selection (individuals' network-related choices) and influence (effect of actors on each other's attributes). The SAOM's are accessible for practitioners through the {\bf RSiena} \citep{RSiena} software package. 

In addition to the continuous-time Markov models, exponential-family random graph models (ERGMs) provide a uniquely different view of social networks. \citet{holland1981exponential} introduce the first exponential-family of probability distributions for directed graphs which are applicable only to cross-sectional social networks. However, \citet{robins2001random} naturally extend this framework by allowing for dependence between graphs across discrete time steps. Moreover, \citet{hanneke2010discrete} define a discrete {\it Temporal ERGM} (TERGM) which assumes an exponential-family model for the transitions between graphs. \citet{krivitsky2013separable} further specify TERGMs with the {\it Separable Temporal ERGM} (STERGM) by postulating that the processes by which actors form and dissolve ties are independent or {\it separable} conditional on the previous state of the network. Various other discrete-time models for social network dynamics provide means for data driven analyses of social systems but drawbacks persist. 

Observed social networks are typically represented by directed (or undirected) graphs where edges indicate the presence of a relationship, e.g., friendship. As a result, complex relations are reduced to a binary indicator. Advances in latent space models for rank data \citep{gormley2007latent} provide new context for conceptualizing this information. \citet{hoff2002latent} summarize general latent space approaches to social network analysis while \citet{handcock2007model} describe an unobserved Euclidean {\it social space} where the actors' locations arise stochastically from a mixture of distributions corresponding to different clusters. These strategies are appealing for their flexibility and interpretability but have only been developed for cross-sectional networks. 

Since latent space approaches to social network analysis postulate the existence of an unobserved space where points represent actors, a natural extension would be to propose a spatial-temporal point process for the underlying dynamics. A major drawback in the current models for social network evolution is the assumption that the set of actors remains fixed over time. In real social systems, e.g., an urban high school, the set of actors is constantly changing so this assumption can be problematic. Spatial birth-death processes \citep{moller2003statistical} offer a stochastic framework for the positions of actors as they enter or exit the system over time. Unfortunately, these process cannot model changes in persistent (present at several consecutive time points) actors' positions. Hence, we seek a stochastic model that can reasonably describe the positions of actors as they enter, navigate, and exit the social space. 

In Section 2, we formally define the social space and derive a discrete-time Markov process to describe fundamental social phenomena. In Section 3, we present analytic results, develop likelihood-based inferential methods, and discuss computation. In Section 4, we apply the methodology to a longitudinal study of adolescent students to explore changes in risky behavior in the context of friendship networks. In Section 5, we discuss the relevance of this work in broader social science research and consider extensions to the modeling framework. 

\section{Point Process Models for Social Systems}

\subsection{Conceptualization}

To conceptualize the methods presented here, consider the population of people in a fixed location over time, e.g., students at an urban middle school. We want to understand the social and behavior dynamics of these people over time. For example, we might ask how a student's social ties  affect her propensity to drink alcohol or engage in risky sex. To do so, we need a rich representation of the time sensitive social landscape. Note that this approach is distinctly different from a traditional panel survey where we attempt to follow a fixed cohort over time. Instead, we focus on the interactions of a dynamic population in a fixed location where we may observe significant composition change within the group between waves of data collection. That is, we do not expect to observe the same set of people at every wave.

Generally, consider a set of actors in social space at time $t$. In the example above, the actors are students and the social space is the school where they interact. Also consider the positions of actors in social space at time $t$. While we formalize this below, the intuition is straightforward: the set of distances between positions in social space represent social relations. For example, two people who have been friends for years tend to be very close to one another in the space whereas casual acquaintances tend to be considerably further apart. 

The major advantage of this conceptualization is flexibility. Complex and nuanced relationships can be accurately represented by a distance metric. Conventionally, we study social networks where relationships are binary, e.g., 1 indicates a friendship nomination and 0 the absence of such a nomination. 

\subsection{Specification}

For $t=0, 1, \dots $, let $N^t = \{1, \dots, n^t\}$ be the set of unique actor labels up to time $t$ with $N^0 \subseteq N^1 \subseteq \cdots$ and let $S^t \subseteq N^t$ denote the set of actors present at time $t$. Further, let $(\mathscr{S}, \norm{\cdot})$ be a normed space where $Z^t = \{Z_i^t \in \mathscr{S}: i \in S^t \}$ is the set of actor locations at time $t$ and $X^t$ is an $n^t \times q$ matrix of actor covariates. We say that $\{S^t, X^t, Z^t\}_{t\geq 0}$ defines a {\it social space}. Next, suppose that $S^t$ (and implicitly $N^t$), $X^t$, and $Z^t$ are random variables that jointly form a stochastic process. If $\{S^t, X^t, Z^t\}_{t\geq 0}$ satisfies the Markov property in time and the transition probability $P(S^t, X^t, Z^t | S^{t-1}, X^{t-1}, Z^{t-1})$ is an exponential family, then we call $\{S^t, X^t, Z^t\}_{t\geq 0}$ a {\it Spatial Temporal Exponential-Family Point Process} (STEPP). Next, we construct a fundamental class of STEPPs by making a few assumptions about the social space and deriving transition distributions. 

{\bf Assumption 1:} $\{S^t\}_{t\geq 0}$ is a process exogenous to $(X^t, Z^t)$. Recall, $S^t$ is the set of actors who are currently in the system at time $t$, e.g., students in a classroom. While one can imagine many scenarios in which the actors who enter or exit the social space is endogenous, e.g., delinquent students are more likely to be expelled, we focus here on the exogenous cases. 

{\bf Assumption 2:} Actor positions in social space, $\{Z_i^t: i \in S^t\}$, are conditionally independent given the previous positions, $Z^{t-1}$. This assumes that actors move through the social space based on the information available at the current time. 

Assumption 1 makes modeling the composition change of actor sets between waves distinctly separate from the changes in actor positions and their corresponding covariates. We refer to $\{S^t\}_{t\geq 0}$ as a {\it migration process} where the actors who enter the system are {\it immigrants} and the actors who exit the system are {\it emigrants}. 
Assumption 2 implies that we can marginalize the transition distributions at the actor-level. Thus, we derive a general class of STEPPs below by specifying the form of
\[
P(Z_i^t, X_i^t | Z^{t-1}, X^{t-1}, S^{t-1})
\]
where it is implicit that $i \in S^t$. We refer to this as the {\it ego transition distribution} (ETD) and it is specified by a series of increasingly complex processes. These processes are basic drift, atomic drift, homophilous attraction, homophilous repulsion, heterophilous attraction, and heterophilous repulsion. 

A {\it basic drift process} describes actor positions only and is determined by a single parameter $\delta_0 \geq 0$. The ETD is given by 
\begin{align}
P_{\delta_0} (Z_i^t | Z^{t-1}, S^{t-1}) = \frac{\exp \left( -\delta_0 \norm{Z_i^t - Z_i^{t-1}} \right) } {c(\delta_0)}
\end{align}
where 
\[
c(\delta_0) = \int_{\mathscr{S}} \exp \left( -\delta_0 \norm{z - Z_i^{t-1}} \right) \cdot \mu(dz)
\]
is the normalizing constant. Note that given the space $(\mathscr{S}, \norm{\cdot})$, the underlying measure $\mu$ must be chosen to ensure $c(\delta_0) < \infty$. A basic drift process is the simplest stochastic model for actor mobility in social space. Along these lines, we also have {\it behavior persistence}. For $m=1, \dots ,q$ and for every $i \in S^{t-1} \cap S^t$, let
\begin{align}
\rho_m = P(X_{im}^t = x | X_{im}^{t-1} = x)
\end{align}
denote the probability that behavior $m$ persists through a single transition. Note that this alone does not completely specify a probability distribution except in the case of a Bernoulli random variable. Also, note that in the case where a covariate is structurally non-random, we can set $\rho_m = 1$. 

To derive more complex processes, we need to formalize the notion of closeness in social space. For any $z \in \mathscr{S}$ and $k \in \mathbb{N}$, consider a set $E \subset \mathscr{S}$ with $|E| < \infty$ and $z \notin E$, where $| \cdot |$ denotes the counting measure. Let
\[
I_1 = \underset{z' \in E}{\text{arg min}} \norm{z-z'}.
\]
For $j=2, \dots, k$, let $J_{j-1} = E \setminus \bigcup_{l=1}^{j-1} I_l$ where
\[
I_j = \underset{z' \in J_{j-1}}{\text{arg min}} \norm{z-z'}.
\]
Then we say that 
\begin{align}
\mathcal{B}_k (z, E) = \bigcup_{j=1}^k I_j
\end{align}
defines a {\it neighbor set} for $z$ where $E$ is the defining expression. Next, let 
$w: \mathscr{S} \times \mathscr{S} \rightarrow [0,1]$ be a weighting function for two positions in social space. If $w$ satisfies 
\advance\parskip by -3pt
\begin{enumerate}[(i)]
\item $w(z, z) = 1$;
\item there exists a $z' \neq z$ such that $w(z, z') = 1$;
\item $w(z, z') \rightarrow 0 \text{ as } \norm{z-z'} \rightarrow \infty$;
\item $w(z, z') \rightarrow 0 \text{ as } \norm{z-z'} \rightarrow 0$.
\end{enumerate}
\advance\parskip by 3pt
then we say that it is an {\it atomic weighting}. For motivation of this definition, see Section 2.3 below. 

Similar to a basic drift process, an {\it atomic drift process} describes actor positions only and is determined by a single parameter $\delta_1 \geq 0$. However, the ETD is considerably more complex. For an atomic weighting $w$, atomic drift is defined by
\begin{align}
P_{\delta_1} (Z_i^t | Z^{t-1}, S^{t-1}) = \frac{ \exp \left( -\delta_1 \sum_{j\in S^{t-1}} {\bf 1} (Z_j^{t-1} \in \mathcal{B}_k (Z_i^{t-1}, Z_{-i}^{t-1} )) w(Z_i^{t-1}, Z_j^{t-1}) \norm{Z_i^t - Z_j^{t-1}} \right) }
{c(\delta_1)}
\end{align}
where $Z_{-i}^{t-1} = Z^{t-1} \setminus \{Z_i^{t-1}\}$ and ${\bf 1} (\cdot)$ is the indicator function. As specified above, $\mathcal{B}_k (Z_i^{t-1}, Z_{-i}^{t-1} )$ is, with some exceptions, the set of $k$ nearest neighbors of ego $i$ at time $t-1$. In the event that $|S^{t-1}| \leq k$, this neighbor set will have fewer than $k$ members and in the event that multiple actors occupy the exact same position at $t-1$, it could have more than $k$ members. Nonetheless, we refer to this as the set of $k$ nearest neighbors for ego $i$ at time $t-1$. Finally, we combine basic drift and atomic drift to define the {\it general drift process} which we denote
\begin{align*}
P_{\delta} (Z_i^t |Z^{t-1}, S^{t-1}) = P_{\delta_0} (Z_i^t |Z^{t-1}, S^{t-1}) P_{\delta_1} (Z_i^t |Z^{t-1}, S^{t-1})
\end{align*}

Next, we introduce homophilous and heterophilous attraction processes.  For a discrete covariate $X_m^t$ and ego $i$, let 
\[
A_{im}^t = \{Z_l^{t-1} \in Z_{-i}^{t-1} : l \in S^{t-1}, X_{im}^t = X_{lm}^{t-1} \}
\]
and
\[
U_{im}^t =  \{Z_l^{t-1} \in Z_{-i}^{t-1} : l \in S^{t-1}, X_{im}^t \neq X_{lm}^{t-1} \}.
\]
Note that natural extensions exist for continuous covariates but we do not explicitly define them here. For the sake of this construction, assume that all covariates are discrete. Given a set of parameters $\alpha_1, \dots, \alpha_q \geq 0$ and an atomic weighting $w$, which we write $w_{ij}^{t-1} = w(Z_i^{t-1}, Z_j^{t-1})$ for simplicity, the ETD of a {\it homophilous attraction process} on the $m$th covariate is
\begin{align}
P_{\alpha_m}(Z_i^t, X_i^t | Z^{t-1}, X^{t-1}, S^{t-1}) = \frac{ \exp \left( -\alpha_m \sum_{j\in S^{t-1}} {\bf 1} (Z_j^{t-1} \in \mathcal{B}_k (Z_i^{t-1},  A_{im}^t)) w_{ij}^{t-1} \norm{Z_i^t - Z_j^{t-1}} \right) }
{c(\alpha_m)}.
\end{align}
The ETD for homophilous attraction on all covariates is defined by
\begin{align}
P_{\alpha}(Z_i^t, X_i^t | Z^{t-1}, X^{t-1}, S^{t-1}) &= \prod_{m=1}^q P_{\alpha_m}(Z_i^t, X_i^t | Z^{t-1}, X^{t-1}, S^{t-1}) \\
& \propto \exp \left( - \sum_{m=1}^q \sum_{j\in S^{t-1}} \alpha_m {\bf 1} (Z_j^{t-1} \in \mathcal{B}_k (Z_i^{t-1},  A_{im}^t)) w_{ij}^{t-1} \norm{Z_i^t - Z_j^{t-1}} \right).
\end{align}
Here, we omit the normalizing constant in the definition and use the proportional symbol, $\propto$. Given a set of parameters $\upsilon_1, \dots, \upsilon_q \geq 0$ and an atomic weighting $w$, the ETD of a {\it heterophilous attraction process} on the $m$th covariate is
 \begin{align}
P_{\upsilon_m}(Z_i^t, X_i^t | Z^{t-1}, X^{t-1}, S^{t-1}) = \frac{ \exp \left( -\upsilon_m \sum_{j\in S^{t-1}} {\bf 1} (Z_j^{t-1} \in \mathcal{B}_k (Z_i^{t-1},  U_{im}^t)) w_{ij}^{t-1} \norm{Z_i^t - Z_j^{t-1}} \right) }
{c(\upsilon_m)},
\end{align}
which is similar to homophilous attraction but with the neighbor set $U_{im}^t$. Naturally, the ETD for heterophilous attraction on all covariates is defined by
\begin{align}
P_{\upsilon}(Z_i^t, X_i^t | Z^{t-1}, X^{t-1}, S^{t-1}) &= \prod_{m=1}^q P_{\upsilon_m}(Z_i^t, X_i^t | Z^{t-1}, X^{t-1}, S^{t-1}) \\
& \propto \exp \left( - \sum_{m=1}^q \sum_{j\in S^{t-1}} \upsilon_m {\bf 1} (Z_j^{t-1} \in \mathcal{B}_k (Z_i^{t-1},  U_{im}^t)) w_{ij}^{t-1} \norm{Z_i^t - Z_j^{t-1}} \right).
\end{align}

Last, we introduce homophilous and heterophilous repulsion. If $(\mathscr{S}, \norm{\cdot})$ is a linear space, we can alter the ETD for attraction to obtain an opposing effect which we refer to as repulsion. Given the determining parameters $\tilde{\alpha}_1, \dots, \tilde{\alpha}_q \geq 0$, the ETD of a {\it homophilous repulsion process} is given by
\begin{eqnarray}
&~&\phantom{\propto\ } P_{\tilde{\alpha}}(Z_i^t, X_i^t | Z^{t-1}, X^{t-1}, S^{t-1})\\ \nonumber
&~&\propto \exp \left( - \sum_{m=1}^q \sum_{j\in S^{t-1}} \tilde{\alpha}_m {\bf 1} (Z_j^{t-1} \in \mathcal{B}_k (Z_i^{t-1},  A_{im}^t)) w_{ij}^{t-1} \norm{Z_i^t - (2Z_i^{t-1} - Z_j^{t-1})} \right).
\end{eqnarray}
Note that repulsion-like distributions are possible in non-linear spaces but are not addressed here. Homophilous repulsion is structurally very similar to homophilous attraction except we replace $\norm{Z_i^t - Z_j^{t-1}}$ with $\norm{Z_i^t - (2 Z_i^{t-1} - Z_j^{t-1})}$ in the ETD. In a linear space, this has the effect of reflecting the point $Z_j^{t-1}$ through $Z_i^{t-1}$ and considering the attraction toward the reflected point which can be viewed as a repulsion away from the original point $Z_j^{t-1}$. Similarly, for parameters $\tilde{\upsilon}_1, \dots, \tilde{\upsilon}_q$, the ETD of a {\it heterophilous repulsion process} is given by
\begin{eqnarray}
&~&\phantom{\propto\ } P_{\tilde{\upsilon}}(Z_i^t, X_i^t | Z^{t-1}, X^{t-1}, S^{t-1})\\ \nonumber
&~&\propto \exp \left( - \sum_{m=1}^q \sum_{j\in S^{t-1}} \tilde{\upsilon}_m {\bf 1} (Z_j^{t-1} \in \mathcal{B}_k (Z_i^{t-1},  U_{im}^t)) w_{ij}^{t-1} \norm{Z_i^t - (2Z_i^{t-1} - Z_j^{t-1})} \right).
\end{eqnarray}

The complete specification for this class of STEPPs is a combination of the processes derived above and an exponential-family model for $P_{\lambda}(S^t |S^{t-1})$ where $\lambda$ is a parameter vector that determines the distribution. Recall that we assume an exogenous migration process which may take many forms, e.g., the number of emigrants follows a binomial distribution and the number of immigrants a Poisson distribution. To preserve generality, we do not further specify this distribution. For homophily (heterophily), either attraction or repulsion can be used but not both simultaneously. Assuming homophilous and heterophilous attraction, we let 
\[
\theta = (\delta_0, \delta_1, \rho_1, \dots, \rho_q, \alpha_1, \dots, \alpha_q, \upsilon_1, \dots, \upsilon_q, \lambda^{\top})^{\top}
\]
denote the complete parameter vector for this class of STEPPs. The complete transition probability is given by 
\begin{align}
P_{\theta} (S^t, X^t, Z^t &| S^{t-1}, X^{t-1}, Z^{t-1}) = P_{\lambda} (S^t | S^{t-1}) \prod_{i \in S^t} P_{\delta} (Z_i^t | Z^{t-1}, S^{t-1}) \\
& \times \exp \left( \sum_{m=1}^q {\bf 1} (X_{im}^t = X_{im}^{t-1}) \log \rho_m + {\bf 1} (X_{im}^t \neq X_{im}^{t-1}) \log (1-\rho_m) \right)  \\
&\times P_{\alpha} (Z_i^t, X_i^t | Z^{t-1}, X^{t-1}, S^{t-1}) P_{\upsilon} (Z_i^t, X_i^t | Z^{t-1}, X^{t-1}, S^{t-1}),
\end{align}
which is an exponential family. Although many other specifications exist for STEPPs, when we write $(S^t, X^t, Z^t) \sim$ STEPP$(\theta)$ it is in reference to this particular class. 

\subsection{Description}

In this section, we further describe and interpret the class of STEPPs constructed above. The previous section provides a formal specification. This section expands the intuition and motivation for each individual process as well as a complete view of the entire model class. 

The drift processes should be regarded as foundational elements for this class of STEPPs. Basic drift is governed by $\delta_0$, a parameter that dictates the magnitude of actors' movements between transitions, and has the simplest ETD. The probability mass in the ETD is symmetric about the ego's previous position and the rate of decay is proportional to $\delta_0$. That is, larger values of $\delta_0$ place more mass near the previous position than would smaller values. The mode of the ETD is always the ego's previous position so basic drift reinforces the notion that actors tend to navigate the social space with respect to their current position rather than jump around sporadically. A STEPP with basic drift alone results in actors generally drifting around the space making predictable, symmetric movements between transitions. 

The ETD of atomic drift is considerably more complex than that of basic drift, but this is essential for ensuring that a specification resembles actual social processes. In essence, the atomic drift process allows other actors to impact the movement of the ego through a transition with the caveat that only a fixed number of them may have an actual effect and their distance relative to the ego largely determines the magnitude of said effect. We use neighbor sets to fix the number of actors in the social space who may have an effect on the ego because it's impractical to assume that the ego is affected by every other actor at a given time. For example, if the social space is a large corporate office with thousands of employees, any one person cannot possibly know everyone else let alone be significantly influenced by them socially. It is more likely that an employee is aware of a few hundred others and noticeably influenced by one or two dozen of them. Thus, we only sum over the $k$ nearest neighbors in the atomic drift ETD. Focusing on the effect of a single neighbor $j$ on the ego $i$, the functional form would be 
\[
\exp \left( - \delta_1 w(Z_i^{t-1}, Z_j^{t-1}) \norm{Z_i^t - Z_j^{t-1}} \right).
\]
This is strikingly similar to the ETD for basic drift with the inclusion of a weight. This is where using atomic weights is crucial. 

Newton's law of universal gravitation tells us that any two bodies will attract one another with a force that is inversely proportional to the square of the distance between them. In particle physics, this force is considered negligible due to the fact that individual atomic masses are extremely small in comparison to surrounding bodies, e.g., the Earth. However, there is a repulsive electromagnetic force between two atoms when the distance between them is small. This force exists due to the negative charge of the electrons associated with each atom. One can imagine a universe where there are no large bodies to dwarf the mass of individual atoms so these forces can coexist. The observable result would be a weak attractive force between atoms that increases as the distance between them decreases. Once the distance becomes sufficiently small, there is a weak repulsive force that increases as the distance between the atoms decreases. Thus, a natural balance arises. 

\citet{schopenhauer1974parerga} cleverly describes this as the {\it porcupine dilemma}: ``a number of porcupines huddled together for warmth on a cold day in winter; but, as they began to prick one another with their quills, they were obliged to disperse. However the cold drove them together again, when just the same thing happened. At last, after many turns of huddling and dispersing, they discovered that they would be best off by remaining at a little distance from one another. In the same way the need of society drives the human porcupines together, only to be mutually repelled by the many prickly and disagreeable qualities of their nature."

As such, we incorporate atomic weights in the ETD for an atomic drift process to provide general attraction between actors while providing stability in the social space over time. In the complete ETD for atomic drift, we combine the effects of each properly weighted neighbor and scale the overall effect by $\delta_1$. Intuitively, the nearest neighbors have the largest effects and the furthest neighbors have the smallest effects except in cases when near neighbors are too close to the ego. Recall that we only require atomic weights to approach zero in the respective limits so the specific functional form may dramatically impact the dynamics of a social space. 

Homophilous and heterophilous attraction are similar to atomic drift but the primary difference is in the specification of neighbor sets. In homophilous (heterophilous) attraction, the set $A_{im}^t$ ($U_{im}^t$) is constructed based on the random state of $X_{im}^t$ which provides a crucial dependence between the ego's social position and behavior. Given the random state of $X_{im}^t$, we consider the set of nearest homophilous (heterophilous) neighbors based on the behavior of those neighbors at time $t-1$ in order to compute the ETD. That is, the ego does not speculate about the future behavior of others. 

Homophilous (heterophilous) repulsion is similar to attraction since we use the same neighbor set $A_{im}^t$ ($U_{im}^t$) in the ETD but the position adjustment is fundamentally different. Recall that for repulsion, we replace $\norm{Z_i^t - Z_j^{t-1}}$ with $\norm{Z_i^t - (2 Z_i^{t-1} - Z_j^{t-1})}$. In the ETD, the term $\norm{Z_i^t - Z_j^{t-1}}$ places some mass of the distribution centered around the position of actor $j$ at time $t-1$. It follows that the term $\norm{Z_i^t - (2 Z_i^{t-1} - Z_j^{t-1})}$ places the same mass centered around the position $2 Z_i^{t-1} - Z_j^{t-1}$. In a linear space, this point is equivalent to the reflection of $Z_j^{t-1}$ through $Z_i^{t-1}$. In this form, it is clear that repulsion is actually an opposing attraction. 

In this model class, each process is straightforward and motivated by basic social forces. As a result, it may be difficult to grasp the gravity of a complete specification. Since we cannot include attraction and repulsion on the same covariate, consider a STEPP with (basic and atomic) drift, homophilous attraction on each covariate and heterophilous repulsion on each covariate. This complete process is extremely complex in its raw functional form but at the core, the ETD has a summation over different effects from neighboring actors to the ego. Each effect is slightly different depending on time-sensitive information (relative distances between actors and behavior) and global properties determined by each parameter. By construction, each parameter is non-negative so we can focus on their relative differences for interpretation. For example, the largest of $\alpha_1, \dots, \alpha_q$ indicates the covariate which exhibits the strongest attraction between similar actors. Alternatively, one of the $\tilde{\upsilon}_1, \dots, \tilde{\upsilon}_q$ being very small or 0 indicates a covariate which exhibits little to no repulsion between dissimilar actors. It is crucial to note that these parameters determine global properties of the social space as opposed to time dependent or individual properties which are the focus of future work. Simulated examples of various STEPPs are available at \url{http://tinyurl.com/STEPPMODELS}.

\section{Statistical Inference}

\subsection{Analysis for a general Euclidean social space}

By slightly restricting the general STEPP model of Section 2, we can derive closed form ETDs and inferential methods. In this section, we show that the ETD for $Z_i^t$ conditional on $X_i^t$ for any subset of the processes described above is multivariate normal if $\mathscr{S} = \mathbb{R}^d$ and the norm is Euclidean distance squared, i.e., $\norm{z} = \sum_{i=1}^d z_i^2$. Based on this result, we derive the marginal ETD for $X_i^t$ and provide a closed form distribution for this class of STEPP models. 

First, assume that $\mathscr{S} = \mathbb{R}^d$ and for $z \in \mathscr{S}$, $\norm{z} = \sum_{i=1}^d z_i^2$. Using a general Euclidean space is somewhat restrictive in a mathematical sense but practically, it provides a flexible, comprehensible foundation for the social space. From this point on, when we say "distance" it is in reference to standard Euclidean distance whereas the norm is specified above. To motivate using the square of Euclidean distance for a norm, we appeal to physics and the inverse square law which generally states 
\[
\text{Intensity} \propto \frac{1} {\text{distance}^2}. 
\]
In practice, we use the atomic weighting function 
\[
w(z_1, z_2) = \begin{cases}
1 &\text{ if } z_1 = z_2 \\
\norm{z_1-z_2} &\text{ if } \norm{z_1-z_2} < c \text{ and } z_1 \neq z_2\\
\norm{z_1-z_2}^{-1} &\text{ if } \norm{z_1-z_2} \geq c,
\end{cases}
\]
where $0 < c \leq 1$ is some threshold. Then when the distance between actors exceeds $\sqrt{c}$, the effect on the ETD is inversely proportional to said distance squared. For shorter distances, we cannot apply the same relationship because it leads to instability as previously discussed. Since many physical phenomena, e.g., Newton's law of universal gravitation, follow an inverse square law, it provides a natural foundation for a Euclidean social space. It is important to note that we need not specify an atomic weighting for the results in the section to hold, but it is necessary to properly motivate this specification. 

Next, we will prove that the ETD has an analytic closed form through a series of lemmas leading up to the final theorem. First, we adopt some notation. For functions $h, g: \mathbb{R}^d \rightarrow \mathbb{R}$, if $h(z) = g(z) + c_0$ where $c_0$ is a constant, we say $h(z) \asymp g(z)$. 

\noindent \underline{Lemma 1:} For $w_1, \dots, w_n \geq 0$ and $\mu_1, \dots, \mu_n \in \mathbb{R}^d$, 
\[
\sum_{j=1}^n w_j \norm{z-\mu_j} \asymp w^* \norm{z-\mu^*/w^*}
\]
where
\[
w^* = \sum_{j=1}^n w_j \hspace{1cm} \text{and} \hspace{1cm} \mu^* = \sum_{j=1}^n w_j \mu_j. 
\]

\begin{proof}
See the Supplementary Materials.
\end{proof}

\noindent \underline{Lemma 2:} Let $Z \in \mathbb{R}^d$ be a random vector with $\mu_1, \dots, \mu_n \in \mathbb{R}^d$, and $w_1, \dots, w_n \geq 0$ where $w^* = \sum_{i=1}^n w_i > 0$ and $ \mu^* = \sum_{i=1}^n w_i \mu_i$ If $P(Z=z) \propto \exp \left\{ -\sum_{i=1}^n w_i \norm{z - \mu_i} \right\}$, then 
\[
Z \sim \mathcal{MVN} \left (\frac{\mu^*}{w^*}, \frac{1}{2w^*} I_d \right ).
\]

\begin{proof}
See the Supplementary Materials.
\end{proof}

\noindent \underline{Theorem:} For each $i \in S^t$, $[Z_i^t |X_i^t, Z^{t-1}, X^{t-1}, S^{t-1}] \sim \mathcal{MVN} (\mu_i^t, \Sigma_i^t)$ where 
\begin{align*}
\mu_i^t = \frac{ \sum_j \theta^{\top} H_{ij}^t w_{ij}^t Z_j^{t-1}} 
{\sum_j \theta^{\top} H_{ij}^t w_{ij}^t}
\hspace*{1cm}
\Sigma_i^t = \left( \frac{1} 
{2 \sum_j \theta^{\top} H_{ij}^t w_{ij}^t} \right) \bf{I_d}
\end{align*}
and an explicit expression for the elements of $H_{ij}^t$ is given 
in the Supplementary Materials. 
\begin{proof}
See the Supplementary Materials.
\end{proof}

Since we assume that the covariates are discrete, it is straightforward to calculate the marginal distribution of $[X_i^t | Z^{t-1}, X^{t-1}]$ based on the theorem. We know $P_{\theta}(Z_i^t, X_i^t | Z^{t-1}, X^{t-1})$ up to a normalizing constant and $P_{\theta}(Z_i^t, | X_i^t, Z^{t-1}, X^{t-1})$ completely so it is possible to integrate out $Z_i^t$ for each value of $X_i^t$. Since the marginal distribution of $Z_i^t$ is multivariate normal, this integral is a function of the variance $\Sigma_i^t$ and the value of $X_i^t$. Then we know $P(X_i^t |Z^{t-1}, X^{t-1})$ up to a normalizing constant for every value of $X_i^t$ and can renormalize these values to obtain the complete distribution. Thus, the complete ETD can be written in closed form as
\[
P_{\theta}(Z_i^t, X_i^t | Z^{t-1}, X^{t-1}) = P_{\theta}(Z_i^t | X_i^t, Z^{t-1}, X^{t-1})  P_{\theta}(X_i^t | Z^{t-1}, X^{t-1}).
\]
Recall that the migration process $\{S^t\}_{t\geq 0}$ is an exogenous exponential family so we have the necessary components for a complete, closed form likelihood. 

\subsection{Likelihood-Based Inference}

In this section, we use the analytic results from the previous section to develop a likelihood-based inferential framework. At this juncture, it is also possible to develop a full Bayesian inferential framework, however here we focus on a Frequentist inference framework. This provides straightforward calculations of parameter estimates and standard errors that might otherwise be computational complex and conceptually challenging to interpret. 
The natural extension to a full Bayesian framework will appear elsewhere.

Suppose that $(S^t, X^t, Z^t) \sim$ STEPP$(\theta)$ for $t=0, \dots, \tau$. That is, this is one STEPP with $\tau$ transitions. For brevity, we suppress the superscripts and simply write $(S, X, Z)$ to denote the complete data over all time steps. Then the likelihood is given by
\begin{align*}
L(\theta | S, X, Z) &= \prod_{t=1}^{\tau} P_{\theta} (S^t, X^t, Z^t | S^{t-1}, X^{t-1}, Z^{t-1}) \\
&= \prod_{t=1}^{\tau} \left( \prod_{i \in S^t} P_{\theta} (Z_i^t, X_i^t | Z^{t-1}, X^{t-1}, S^t) \right) P_{\theta} (S^t | S^{t-1}) \\
&= \prod_{t=1}^{\tau} \left( \prod_{i \in S^t} P_{\theta} (Z_i^t | X_i^t, Z^{t-1}, X^{t-1}, S^t) P_{\theta} (X_i^t | Z^{t-1}, X^{t-1}, S^{t}) \right) P_{\theta} (S^t | S^{t-1}).
\end{align*}
It is implicit in this formulation that the initial state $(S^0, X^0, Z^0)$ is fixed and not random. It is natural to extend this model class to allow for a random initial state but it is not explored here. However, we must note that the parameters in this class of STEPPs determine transitions between states rather than isolated states so a model for $(S^0, X^0, Z^0)$ may be difficult to align conceptually. 

As shown above, the likelihood function has a computationally closed form so we can use standard optimization routines to obtain parameter estimates and standard errors. However, calculating the likelihood can be cumbersome due to the inherent complexity of each ETD. In the next section, we address these issues and provide a general computational framework for performing likelihood-based inference. 

\subsection{Computation}

In this section, we describe the computational challenges of implementing likelihood-based inference for STEPP data. The likelihood function provided above is straightforward to calculate but doing so may be computationally expensive. Since the migration process $\{S^t\}_{t\geq 0}$ is exogenous, we focus on elements of the likelihood that involve actor positions and covariates. Explicitly, we need to calculate
\[
\prod_{t=1}^{\tau} \prod_{i \in S^t} P_{\theta} (Z_i^t | X_i^t, Z^{t-1}, X^{t-1}, S^t) P_{\theta} (X_i^t | Z^{t-1}, X^{t-1}, S^{t}).
\]

As shown previously, $[Z_i^t | X_i^t, Z^{t-1}, X^{t-1}]$ follows a multivariate normal distribution so calculating
\[
P_{\theta} (Z_i^t | X_i^t, Z^{t-1}, X^{t-1}, S^t)
\]
given parameters $\mu_i^t$ and $\Sigma_i^t$ is extremely fast. However, calculating these parameters can be computationally demanding. For each time period $t$ and ego $i\in S^t$, we must compute multiple pairwise distances, weights, neighbor sets and sum over every element. Above all, computing neighbor sets is the most demanding. Given $q$ covariates, a full specification requires computing up to $2q + 1$ neighbor sets for each ego. 

While we have shown that one can calculate the distribution of $[X_i^t | Z^{t-1}, X^{t-1}]$ for an arbitrary discrete covariate, we focus on the case where the support of each is finite. Recall that $X_i^t$ is a random vector with $q$ components and for each of them, we calculate every value in the probability mass function. Each calculation has a closed from but crucially depends on the variance $\Sigma_i^t$ which is computationally demanding. Thus, we must calculate $\Sigma_i^t$ conditional on each element in the full support of the vector $X_i^t$ to obtain $P_{\theta} (X_i^t | Z^{t-1}, X^{t-1}, S^{t})$ as required. 

In practice, we maximize the log likelihood function
\[
\ell (\theta) = \sum_{t=1}^{\tau} \sum_{i \in S^t} \log P_{\theta} (Z_i^t | X_i^t, Z^{t-1}, X^{t-1}, S^t) + \log P_{\theta} (X_i^t | Z^{t-1}, X^{t-1}, S^{t})
\]
since it is slightly more stable numerically. Explicitly, 
\[
\hat{\theta} = \underset{\theta \in \Theta}{\text{ arg max }} \ell (\theta)
\]
is the maximum likelihood estimator. 

\subsection{Analysis}

In this sub-section we consider the properties of the maximum likelihood estimator. The asymptotic properties of the MLE will depend on the framework
the inference is embedded in.
If we could observe a sequence of independent and identically distributed
(IID) STEPPs, standard large sample theory would imply that $\hat{\theta}$
is consistent and asymptotically efficient \citep{casella2002statistical}.
Moreover, one can derive a standard Central Limit Theorem for this
situation. However, in reality, we rarely can observe a
sequence of IID STEPPs since social systems are constantly evolving, and 
we do not detail analytic results pertaining to this case.


\begin{figure}[H]
\centering
\includegraphics[width=0.60\textwidth]{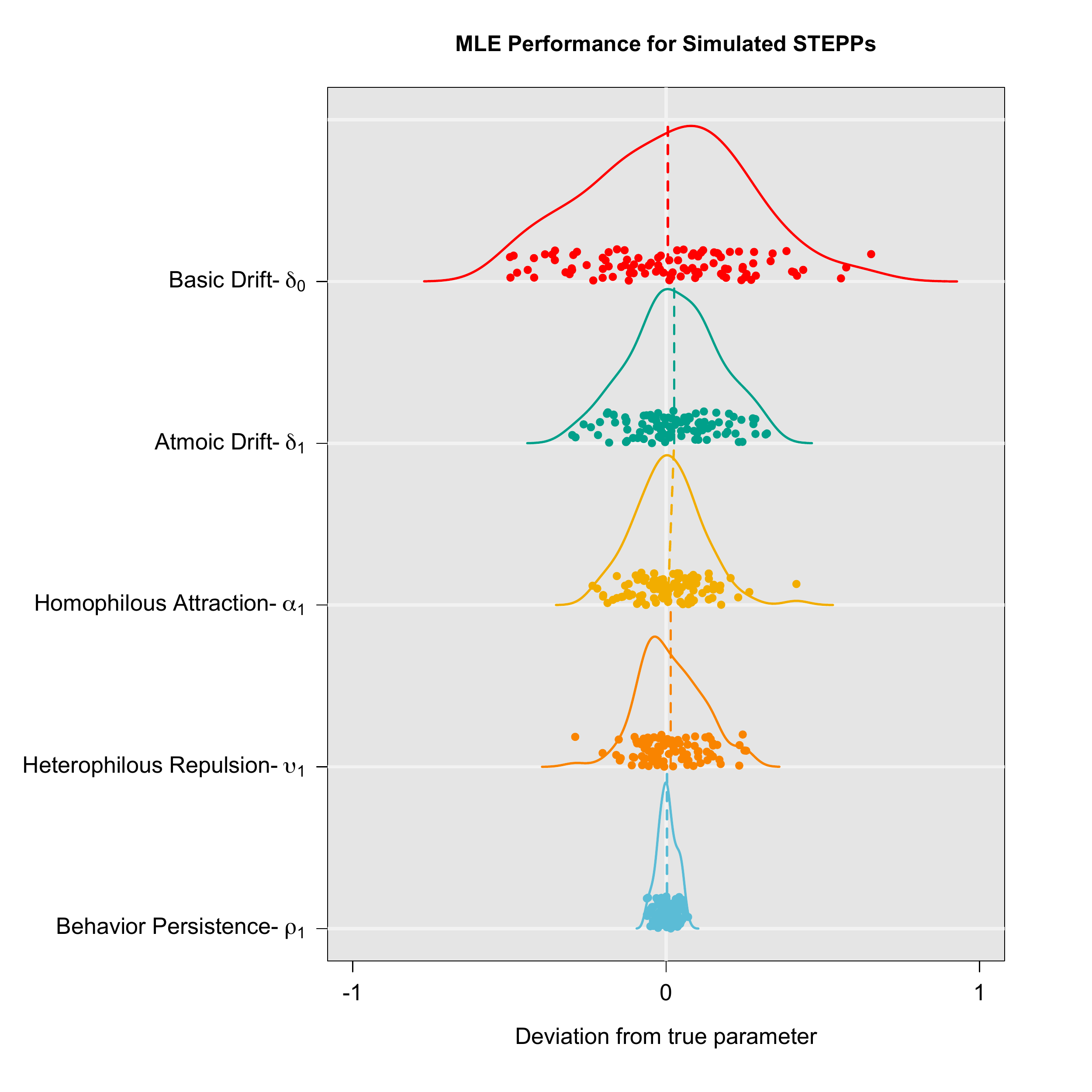}
\caption{Maximum likelihood estimates of parameters over 100 simulated STEPPs with a kernel density of each sample overlayed and sample means marked with dashed lines.}
\end{figure}

Here we explore the properties for the most common situation where a single STEPP is observed through a moderate number of time points. 
Specifically, we generated
100 independent STEPPs each with one binary covariate and 50 actors across
five time periods and with no migration. Further, we used homophilous attraction and
heterophilous repulsion on the single covariate. The parameter values were set
to moderate effects, as specified in the first row of Table 1. 
For each
generated STEPP, we computed the MLE estimator $\hat{\theta}$ and display
estimates of them in Figure 1. Note that the true parameter value is subtracted from each estimate
for comparison, i.e., zero corresponds to the true value. Additionally, we
mark the sample mean of each sample with a vertical bar and overlay a
kernel density estimate for each set of estimates. 
We see that the distribution are centered around the true values and the shapes are approximately Gaussian.

As we have an explicit and computable expression for the log-likelihood, we can employ it to summarize the inference. We can compute standard errors
from the numerical Hessian of $\ell (\theta)$. In Table 1 we summarize the standard errors estimates and compare them to their true values.
\begin{table}[ht]
\centering
\begin{tabular}{r|rrrrr}
  \hline
 & $\delta_0$ & $\delta_1$ & $\alpha_1$ & $\tilde{\upsilon}_1$ & $\rho_1$ \\ 
  \hline
true parameter & 0.50 & 0.50 & 1.00 & 0.75 & 0.80 \\ 
  mean of MLE estimate & 0.52 & 0.52 & 0.99 & 0.75 & 0.80 \\ 
  std. dev. of MLE estimates & 0.29 & 0.24 & 0.12 & 0.11 & 0.03 \\ 
  mean of SE estimates & 0.25 & 0.19 & 0.14 & 0.10 & 0.03 \\ 
  std. dev. of SE estimates & 0.05 & 0.02 & 0.01 & 0.01 & $<0.01$ \\ 
   \hline
\end{tabular}\caption{Assessing the standard error estimates. We simulate 100 STEPPs with 50 actors and 5 time transitions then compute the MLE and standard error for each. This table shows the sample mean and sample standard deviation of the MLE and standard error estimate of each parameter. In all cases, both the MLE and standard error estimates are close to their true values.}
\end{table}
These simulation results support the use of likelihood-based inference and suggest that the MLE and standard errors will be credible. In most
situations where this estimator will be used the amount of information will be at the level of this simulation or higher.

The explicit form of the log-likelihood, $\ell (\theta),$ also make it possible to assess the goodness-of-fit of a
model via an analysis of deviance.  Specifically, we can compute the
change in log-likelihood from the null model $(\theta=0)$ to the
MLE. Similarly, we can the graphical goodness-of-fit ideas to assess the overall fit of the model
\citep{hungoodhan07}. To do so we
can use temporal-structural network summary statistics (e.g., persistence of ties, counts of nodal types) to compare
observed behavior with those produced by the STEPP model.

\section{Application to Adolescent Risk Behavior in Networks}

In this section we apply to a longitudinal network of friendships within a school. The primary question of interest is the coevolution of risky behavior and friendship ties. More specifically, the interaction between social forces and substance use in adolescents has drawn significant attention in recent research. In particular, researchers are interested in quantifying the effect of peers on individual behavior as well as the effect an individual may have on their peers. \citet{brechwald2011beyond} summarize recent advances in the study of peer influence and explore the range of behaviors for which peer influence occurs. \citet{poulin2011short} provide a longitudinal analysis of friendship selection on adolescent cigarette, alcohol, and marijuana use. Their analysis depends on a cross-lag panel model that tests for the reciprocated association between substance use and the number of new friends who use the same substance. While this work provides unique insight into peer influence on adolescent substance use, the lack of sophisticated models for complex social processes is problematic. 

The SAOMs discussed in section 1 have been applied by \citet{delaHaye2013selection} to explicitly model selection and influence in adolescent social networks with respect to marijuana use at two schools in the Add Health study. Their results indicate that having friends who have used marijuana in their lifetime is not a significant predictor of individual initiation while recent (within the last six months) use is a significant predictor of individual initiation at one of the schools. \citet{tucker2014peer} use SAOMs to model selection and influence effects of marijuana more extensively in the Add Health study. They found that in one school, influence occurred in reciprocated relationships which are hypothesized to be characterized by closeness and trust. However, in another school it was found that adopting friends' drug use behavior appears to be a strategy for attaining social status. SAOMs provide researchers with a sophisticated model for beginning to disentangle social forces and behavior, but it requires strong assumptions. Specifically, the set of actors is assumed to remain fixed over time and the coevolution of social structure and behavior is based on friendship tie variables which may unstable. 

\subsection{CARBIN Study}

The Contextualizing Adolescent Risk Behavior In Networks (CARBIN) study was designed to investigate changes in risky behavior, e.g., drug use, amongst middle and high school students in the context of directed friendship networks. While several waves of data were collected from a few urban schools in Peoria, IL, we use four waves of data collected from one school for this illustration. 

Each student student filled out an extensive survey that asked them about their personal behavior and to nominate their friends within the school. We utilize these friendship nominations to construct social networks and focus on two substance use variables: any alcohol and any marijuana use in the last 30 days. Table 2 summarizes the raw data of interest for this exercise. Since there are only three time transitions and the amount of composition change is relatively modest, we do not explore models for the migration process. Instead, we focus on the latent social space and the processes that govern it. In the next section, we fit latent positions to the observed networks and estimate STEPP parameters. 

\begin{table}[ht]
\centering
\begin{tabular}{rrrrr}
  \hline
 & wave 1 & wave 2 & wave 3 & wave 4 \\ 
  \hline
respondents & 150 & 168 & 173 & 168 \\ 
  immigrants & 0 & 30 & 5 & 11 \\ 
  emigrants & 0 & 7 & 19 & 72 \\ 
  alcohol users & 52 & 53 & 60 & 36 \\ 
  marijuana users & 9 & 13 & 13 & 10 \\ 
  friendship ties & 754 & 842 & 707 & 317 \\ 
   \hline
\end{tabular}
\caption{Summary counts for the CARBIN study (by wave)}
\end{table}

\subsection{Implementation and Estimation}

To estimate STEPP parameters, we first need to infer actor positions in the latent social space from observed networks. \citet{krivitsky2008fitting} provide a compelling framework for fitting latent position cluster models to cross-sectional social networks. Based on this work, we use the {\bf latentnet} package \citep{latentnet} to fit latent positions for each wave of data. The standard procedure for fitting such positions uses the minimum Kullack-Leibler (MKL) divergence from a prior distribution which is invariant under rotation and scaling. Since we are concerned with the transitions between time periods, each set of cross-sectional positions should reasonably align with the others. 

The data exhibit strong non-planar patterns in space so we fit positions in three dimension $(d=3)$. First, we fit an aggregate model to a single combination of all four networks to obtain a reference point for each individual cross-section. It is important to note that we only use the observed friendship ties and do not adjust for any covariate information in this process. 
Next, we fit latent positions to each cross-sectional network using this set of points from the aggregate model as references. That is, the reference points provide starting values for the optimization process. Finally, we use Procrustes analysis to minimize any residual variance from rotation or scaling between waves. The final sequence of points is referred to in the standard STEPP notation. Thus, we have actor positions $Z^0, Z^1, Z^2, Z^3$ in three dimensions, covariate matrices $X^0, X^1, X^2, X^3$ where the first column is a binary indicator of individual alcohol use and the second column is a binary indicator of individual marijuana use, and persistence sets $S^0, S^1, S^2, S^3$ that indicate which actors are present at each point in time.

Using the inferential framework presented in Section 3, we estimate a STEPP model for this data. Consider a model with a drift process (basic and atomic), homophilous attraction on both variables, and heterophilous repulsion on both variables. We use homophilous attraction because it is reasonable to assume that students who use alcohol or marijuana are likely to attract other students who use and vice versa. Similarly, we use heterophilous repulsion because it is more likely that students who do not share similar substance use behavior are more likely to be repelled by one another than attracted. 


With a fully specified model, we can state a null hypothesis regarding the parameter values and use the likelihood to compute standard errors and $p-$values. We consider the null hypothesis,
\begin{align*}
\delta_0 &= 1 ~~~~~~~~~~~
\delta_1 = \alpha_1 = \alpha_2 = \tilde{\upsilon}_1 = \tilde{\upsilon}_2 = 0 ~~~~~~~~~~~
\rho_1 = \rho_2 = 0.5.
\end{align*}
That is, we assume that the only process at work is basic drift and the behavior persistence terms are equivalent to a fair coin flip. Note that setting $\delta_0 = 1$ in the null is somewhat arbitrary but the other parameters are of primary interest. To obtain standard errors, we use a standard estimate of the Hessian of the likelihood function and then base nominal $p-$values on the nominal limiting normal distribution discussed in Section 3. In the next section, we report results and provide a brief interpretation. 

\subsection{Results}

The results from the estimation are summarized in Table 3. 
We observe that all of the processes except heterophilous repulsion on marijuana are significant at the 10\% level. 

\begin{table}[H]
\centering
\begin{tabular}{| l | c  c  r |}
\hline
Parameter & Estimate & Std. Error & p-value \\
\hline
Basic drift - $\delta_0$ & 0.0817 & (0.018) & $< 0.0001^{***}$ \\
Atomic drift - $\delta_1$ & 0.0707 & (0.041) &  $0.0873.\kern1pt~~~$\\
Alcohol attraction - $\alpha_1$ & 0.0766 & (0.027) & $0.0053^{**}$\phantom{i}\\
Marijuana attraction - $\alpha_2$ & 0.0984 & (0.042) &  $0.0181^{*~~~}$\\
Alcohol repulsion - $\tilde{\upsilon}_1$ & 0.1084 & (0.018) & $<0.0001^{***}$ \\
Marijuana repulsion - $\tilde{\upsilon}_2$ & 0.0000 & (0.000) & $0.9980$\phantom{i...}\\
Alcohol persistence - $\rho_1$ & 0.7077 & (0.024) & $< 0.0001^{***}$\\
Alcohol persistence - $\rho_2$ & 0.9248 & (0.019) & $< 0.0001^{***}$ \\
\hline
\end{tabular}
\caption{Summary of STEPP model estimates for the CARBIN data.}
\end{table}

To compare the effects of each process on the social system, we rescale the estimated parameters to produce a relative interpretation. Recall that the parameters can be viewed as pseudo-scaling factors in calculating the mean and variance of a normal distribution. That is, scaling all of the parameters would have a small or negligible effect on the mean and an inversely proportional effect on the variance. Intuitively, the mean of each ETD provides information regarding where actors are likely to move and the variance provides information regarding the range of those potential movements. Hence, we report $\delta_0^*, \delta_1^*, \alpha_1^*, \alpha_2^*, \tilde{\upsilon}_1^*,  \tilde{\upsilon}_2^*$ and $\tau = \delta_0 + \delta_1 + \alpha_1 + \alpha_2+ \tilde{\upsilon}_1+  \tilde{\upsilon}_2$ where 
\[
\delta_0^* = \delta_0/\tau, \hspace*{0.5cm}
\delta_1^* = \delta_1/\tau, \hspace*{0.5cm}
\alpha_1^* = \alpha_1/\tau, \hspace*{0.5cm}
\alpha_2^* = \alpha_2/\tau, \hspace*{0.5cm}
\tilde{\upsilon}_1^* = \tilde{\upsilon}_1/\tau, \hspace*{0.5cm}
\tilde{\upsilon}_2^* = \tilde{\upsilon}_2/\tau.
\]
These rescaled parameters reflect the proportion of actors' movements through the social space, e.g., $\delta_0^* = 0.1$ would imply that 10\% of social movement is attributable to basic drift. The rescaled parameters in Table 3 tell a more compelling story about this social space.
\begin{table}[h]
\centering
\begin{tabular}{c | c | c | c | c | c | c }
Basic & Atomic & Alcohol & Marijuana & Alcohol & Marijuana & Sum \\
$\delta_0^*$ &$\delta_1^*$ &$\alpha_1^*$ &$\alpha_2^*$ & $\tilde{\upsilon}_1^*$ & $\tilde{\upsilon}_2^*$ & \\
\hline
0.187 & 0.162 &0.176 &0.226 & 0.249 & 0.000 & 1.000
\end{tabular}

\begin{tabular}{ c | c | c }
Variation & Alcohol & Marijuana \\
$\tau$ & $\rho_1$ & $\rho_2$ \\
\hline
0.436 & 0.708 & 0.925
\end{tabular}
\caption{Rescaled STEPP parameters}
\end{table}
We observe that 18.7\% of movement is basic drift, 16.2\% is atomic drift,
17.6\% is homophilous attraction on alcohol use, 22.6\% is homophilous
attraction on marijuana use, 24.9\% is heterophilous repulsion based on
alcohol use, and 0\% is heterophilous repulsion on marijuana use. The
persistence for alcohol use (or non use) is 70.8\% and the persistence for
marijuana use (or non use) is 92.5\%. 

Using STEPPs to model adolescent substance use provides a new conceptualization and quantification of the social forces at play in a community. Since the complexity of social networks is captured by distance in Euclidean space, we do not need to model nuanced changes in friendship ties or shared substance use. Instead, we simply model the fundamental forces of attraction and repulsion as they pertain to each substance. In the example above, we observe that the forces of homophilous attraction and heterophilous repulsion on alcohol are very strong, accounting for a combined 42.5\% of the actors' movements. Conversely, there is no heterophilous repulsion on marijuana and a modest homophilous attraction contributing to 22.6\% of actors' movements. This result tells a compelling story about the difference between alcohol and marijuana in this school. Based on the model formation, we can infer that shared alcohol use (or non-use) is pushing students together while students with opposing usage are driven apart, and some students are influenced to adopt new behaviors based on those closest to them in social space. Conversely, shared marijuana use (or non-use) is pushing students together while opposing usage has no effect on driving them apart. Although, it must be noted that this is not a statement regarding the general influence of alcohol or marijuana since our analysis is merely an illustration of the methods presented in this paper. It is possible that we are observing the effects of social stigma. That is, marijuana is less prevalent in the example and might be stigmatized compared to alcohol. Formalizing and testing this notion is the topic of future work. 

\section{Discussion}

This paper introduces a novel class of stochastic models for the coevolution of social structure and individual behavior over time. This model class is built on ideas successful applied in latent space approaches to network analysis, longitudinal social network analysis, and spatial-temporal point processes with specific components being motivated by physics and psychology. Additionally, realistic specifications of the broader class lead to traditional likelihood-based inference and computationally tractable solutions to estimation problems. 

As shown in Section 2, complex social systems can be stochastically represented by a set of ego centric processes. The drift process provides an intuitive baseline for the ways in which actors navigate a social space with respect to their own position and neighboring actors' positions. Moreover, the atomic drift process incorporates principles from particle physics to produce stable dynamics over time while Schopenhauer's porcupine dilemma provides a philosophical argument for the functional form of the transition distribution. The attraction and repulsion processes introduce a fundamental dependence between the evolution of individual behavior and one's social position over time. That is, these processes shed light on selection and influence phenomena in a way that is distinctly different from existing frames like the SAOMs. 

While these models reflect real social processes, Section 3 shows that natural specifications also lead to multivariate normal conditional ETDs and computationally tractable likelihood-based inference. It is easy to get lost in the technical details and lose sight of the overarching elegance in these results. Recall that we implement a probabilistic version of the inverse square law, one of the most fundamental relationships in nature, and the result is an analytically closed form transition distribution. Furthermore, that distribution is one of the most fundamental in science, nature and statistical theory: the multivariate normal distribution. Although the parameters of this distribution can be cumbersome to calculate, the result makes likelihood-based inference possible. 

Section 4 illustrates the core methods developed in this paper with an application to a study of alcohol and drug use in adolescent friendship networks. This shows the potential of these methods in numerous applications, but it also highlights future challenges. The process of fitting latent positions to cross-sectional networks and minimizing the variation across observed points in time can be very nuanced and challenging. It is not practical for applied researchers interested in implementing these models to perform this exercise every time. Hence, the focus of future work in this area is to develop more holistic approaches to inferring latent actor positions. Alternatively, this class of models may lead to different forms of data collection in social systems that inform the latent positions more accurately than social ties. 

The STEPP framework for social systems can be extended both methodologically and in application. In this paper, we focus on population level forces but it is natural to extend the model to allow for individual or community level forces. A natural example is to extend the notion of attractive forces and allow actors to have different masses. That is, some actors may have inherently more social influence or attractiveness. Also, we might allow for differential homophily or heterophily, e.g., the attractive force between non-users is weaker than the force between users. In addition to methodological extensions, there are applications of the STEPP framework that do not require solving inferential problems. For example, one can use STEPP simulation as a virtual laboratory for intervention assessment. Given reasonable assumptions regarding actor positions and parameter values, it is possible to stage hypothetical interventions and simulate possible outcomes. Consider a class of 100 students where 40 of them are known binge drinkers and the administration has two options: they can target a few students and conduct intense personal interventions that are 90\% effective or implement a binge drinking prevention program that every student participates in but is only 25\% effective. Since targeting popular students could have spillover effects, it is unclear which option is best. Furthermore, it is unclear which students to target. Simulated STEPPs shed light on an otherwise uncertain decision process. In conclusion, the STEPP framework provides realistic stochastic representations of complex social systems which provides novel tools for social science research.  
 
\section*{Acknowledgements}
\rm
Research reported in this publication release was supported by the National Institute of Drug Abuse of the National Institutes of Health under award number R01DA033280. 
The project described was supported by grant numbers 1R21HD063000 and
5R21HD075714-02 from NICHD, grant number N00014-08-1-1015 from the ONR,
grant numbers MMS-0851555, MMS-1357619 from the NSF.  We are grateful to
the California Center for Population Research at UCLA (CCPR) for general
support. CCPR receives population research infrastructure funding
(R24-HD041022) from the Eunice Kennedy Shriver National Institute of Child
Health and Human Development (NICHD).  Its contents are solely the
responsibility of the authors and do not necessarily represent the official
views of the Demographic \& Behavioral Sciences (DBS) Branch, the National
Science Foundation, the Office of Naval Research, or the National
Institutes of Health. The authors would also like to acknowledge Kayla de la Haye and the principal investigators of the CARBIN study, Harold D. Green, Jr. and Dorothy Espelage, for facilitating this work and motivating the methods presented in this paper. 

\section*{Appendix}
\subsection*{Covariance Matrix in Theorem 1}\label{app:H}
In Theorem 1, the expression for the covariance is:
\[
H_{ij}^t = \left(
\begin{array}{c}
{\bf 1}(j=i) \\
{\bf 1} (Z_j^{t-1} \in \mathcal{B}_k (Z_i^{t-1}, Z_{-i}^{t-1}))\\
0 \\
\vdots \\
0 \\
 {\bf 1} (Z_j^{t-1} \in \mathcal{B}_k (Z_i^{t-1}, A_{i1}^t)) \\
\vdots \\
 {\bf 1} (Z_j^{t-1} \in \mathcal{B}_k (Z_i^{t-1}, A_{iq}^t)) \\
 {\bf 1} (Z_j^{t-1} \in \mathcal{B}_k (Z_i^{t-1}, U_{i1}^t)) \\
\vdots \\
 {\bf 1} (Z_j^{t-1} \in \mathcal{B}_k (Z_i^{t-1}, U_{iq}^t)) \\
0 
\end{array} \right)
\]
where the 0s are matched to $\rho_1, \dots, \rho_q$ and $\lambda$ in the parameter $\theta$. 

\subsection*{Proof of Results}\label{app:sdtergm-terms}
\medskip
In this appendix we provide proofs of the two lemmas and theorem in Section 3.
\subsection*{Proof of Lemma 1}
\medskip

\begin{proof}

\underline{Base case:} $w_1 \norm{z-\mu_1} + w_2 \norm{z-\mu_2} \asymp (w_1 + w_2) \norm{z-(w_1 \mu_1 + w_2 \mu_2)/(w_1 + w_2)}$. 
Initially, the subscripts on $\mu_1$ and $\mu_2$ will be set to superscripts so the subscript can denote individual components. First,  
\begin{align*}
w_1 \norm{z-\mu^1} &= w_1 \sum_{i=1}^d (z_i - \mu_i^1)^2 \\
&= w_1 \sum_{i=1}^d (z_i^2 - 2z_i \mu_i^1 + 2(\mu_i^1)^2 ) \\
&\asymp w_1 \sum_{i=1}^d (z_i^2 - 2z_i \mu_i^1).
\end{align*}
Then 
\begin{align*}
w_1 \norm{z-\mu^1} + w_2 \norm{z-\mu^2} &\asymp w_1 \sum_{i=1}^d (z_i^2 - 2z_i \mu_i^1) + w_2 \sum_{i=1}^d (z_i^2 - 2z_i \mu_i^2) \\
&=\sum_{i=1}^d (w_1 z_i^2 -2w_1 z_i \mu_i^1 + w_2 z_i^2 - 2 w_2 z_i \mu_i^2) \\
&= \sum_{i=1}^d  ((w_1 + w_2) z_i^2 -2 z_i (w_1 \mu_i^1 + w_2 \mu_i^2)) \\
&= (w_1 + w_2) \sum_{i=1}^d (z_i^2 - 2 z_i (w_1 \mu_i^1 + w_2 \mu_i^2)/(w_1 + w_2)) \\
& \asymp (w_1 + w_2) \sum_{i=1}^d (z_i - (w_1 \mu_i^1 + w_2 \mu_i^2)/(w_1 + w_2))^2 \\
&= (w_1 + w_2) \norm{z - (w_1 \mu^1 + w_2 \mu^2)/(w_1 + w_2)}. 
\end{align*}
\underline{Induction step:} Assume $\sum_{j=1}^n w_j \norm{z-\mu_j} \asymp w^* \norm{z-\mu^*/w^*}$ for $n=k$ and show true for $n=k+1$. 
Let $w' = \sum_{j=1}^k w_j$ and $w'' = \sum_{j=1}^{k+1} w_j$. Similarly, let  $\mu' = \sum_{j=1}^k w_j \mu_j$ and $\mu'' = \sum_{j=1}^{k+1} w_j \mu_j$. Then
\begin{align*}
\sum_{j=1}^{k+1} w_j \norm{z-\mu_j} &= \sum_{j=1}^{k} w_j \norm{z-\mu_j} + w_{k+1} \norm{z-\mu_{k+1}} \\
&\asymp w' \norm{z-\mu'/w'} + w_{k+1} \norm{z-\mu_{k+1}} \\
&\asymp w^* \norm{z-\mu^*/w^*}, 
\end{align*}
where
\[
w^* = \sum_{j=1}^k w_j + w_{k+1} = w''
\]
and
\begin{align*}
\mu^* &= w' \frac{\mu'}{w'} + w_{k+1} \mu_{k+1} \\
&= \mu' + w_{k+1} \mu_{k+1} \\
&= \sum_{j=1}^d w_j \mu_j + w_{k+1} \mu_{k+1} \\
&= \mu''
\end{align*}
\end{proof}

\subsection*{Proof of Lemma 2}
\medskip

\begin{proof}
First, observe that if $h(z) \asymp g(z)$, then $e^{h(z)} \propto e^{g(z)}$. Then by Lemma 1, 
\begin{align*}
P(Z=z) &\propto \exp \left\{ -\sum_{i=1}^n w_i \norm{z - \mu_i} \right\} \\
&\propto \exp \left\{ -w^* \norm{z - \frac{\mu^*}{w^*}} \right\}.
\end{align*}
Since we can rewrite $\norm{z} = z^{\top} z$, 
\begin{align*}
P(Z=z) &\propto \exp \left\{ - w^* \left(z - \frac{\mu^*}{w^*} \right)^{\top} \left(z - \frac{\mu^*}{w^*} \right) \right \} \\
&= \exp \left\{ -\frac{1}{2} \left(z - \frac{\mu^*}{w^*} \right)^{\top} \left( \frac{1}{2w^*} I_d \right )^{-1} \left(z - \frac{\mu^*}{w^*} \right) \right \}.
\end{align*}
Therefore,
\[
P(Z=z) = (2\pi)^{-d/2} (2 w^*)^{d/2} \exp \left\{ -\frac{1}{2} \left(z - \frac{\mu^*}{w^*} \right)^{\top} \left( \frac{1}{2w^*} I_d \right )^{-1} \left(z - \frac{\mu^*}{w^*} \right) \right \}.
\]
\end{proof}

\subsection*{Proof of the Theorem}
\medskip

\begin{proof}
First, we need to verify the marginal distribution of $[Z_i^t |X_i^t, Z^{t-1}, X^{t-1}, S^{t-1}]$ up to a normalizing constant. Recall the complete STEPP distribution
\begin{align*}
P_{\theta} (S^t, X^t, Z^t | S^{t-1}, X^{t-1}, Z^{t-1}) &= P_{\lambda} (S^t | S^{t-1}) \prod_{i \in S^t} P_{\delta} (Z_i^t | Z^{t-1}, S^{t-1}) \\
& \times \exp \left( \sum_{m=1}^q {\bf 1} (X_{im}^t = X_{im}^{t-1}) \log \rho_m + {\bf 1} (X_{im}^t \neq X_{im}^{t-1}) \log (1-\rho_m) \right)  \\
&\times P_{\alpha} (Z_i^t, X_i^t | Z^{t-1}, X^{t-1}, S^{t-1}) P_{\upsilon} (Z_i^t, X_i^t | Z^{t-1}, X^{t-1}, S^{t-1}).
\end{align*}
By marginalizing and conditioning on $S^t$ and $X_i^t$, we can reduce this to 
\[
P(Z_i^t | X_i^t, Z^{t-1}, X^{t-1}) = P_{\delta} (Z_i^t | Z^{t-1}, S^{t-1}) P_{\alpha} (Z_i^t, X_i^t | Z^{t-1}, X^{t-1}, S^{t-1}) P_{\upsilon} (Z_i^t, X_i^t | Z^{t-1}, X^{t-1}, S^{t-1}).
\]
Since each term on the right hand side has an exponential form, the exponents sum as follows 
\begin{align*}
&\delta_0 \norm{Z_i^t - Z_i^{t-1}} + \delta_1 \sum_{j \in S^t} {\bf 1} (Z_j^{t-1} \in \mathcal{B}_k (Z_i^{t-1}, Z_{-i}^{t-1})) w_{ij}^{t-1} \norm{Z_i^t - Z_j^{t-1}} \\
&+ \sum_{m=1}^q \sum_{j \in S^t} \alpha_m {\bf 1} (Z_j^{t-1} \in \mathcal{B}_k (Z_i^{t-1}, A_{im}^t)) w_{ij}^{t-1} \norm{Z_i^t - Z_j^{t-1}} \\
&+ \sum_{m=1}^q \sum_{j \in S^t} \upsilon_m {\bf 1} (Z_j^{t-1} \in \mathcal{B}_k (Z_i^{t-1}, U_{im}^t)) w_{ij}^{t-1} \norm{Z_i^t - Z_j^{t-1}} \\
&= \sum_{j \in S^t} (\delta_0 {\bf 1}(i=j) + \delta_1 {\bf 1} (Z_j^{t-1} \in \mathcal{B}_k (Z_i^{t-1}, Z_{-i}^{t-1})) + \sum_{m=1}^q \alpha_m {\bf 1} (Z_j^{t-1} \in \mathcal{B}_k (Z_i^{t-1}, A_{im}^t)) \\
&+ \sum_{m=1}^q \upsilon_m {\bf 1} (Z_j^{t-1} \in \mathcal{B}_k (Z_i^{t-1}, U_{im}^t))) w_{ij}^{t-1} \norm{Z_i^t - Z_j^{t-1}} \\
&= \sum_{j \in S^t} \theta^{\top} H_{ij}^t w_{ij}^t \norm{Z_i^t - Z_j^{t-1}}.
\end{align*}
Hence, 
\[
P(Z_i^t | X_i^t, Z^{t-1}, X^{t-1}) \propto \exp \left(-\sum_{j \in S^t} \theta^{\top} H_{ij}^t w_{ij}^t \norm{Z_i^t - Z_j^{t-1}} \right). 
\]
and Lemma 2 implies that  $[Z_i^t |X_i^t, Z^{t-1}, X^{t-1}, S^{t-1}] \sim \mathcal{MVN} (\mu_i^t, \Sigma_i^t)$.
\end{proof}

\bibliographystyle{plainnat}
\nocite{*}
\bibliography{stepp}

\end{document}